\documentclass[11pt,dvips]{article}
\usepackage{epsfig,times} 
\usepackage{picinpar}
\usepackage{wrapfig}
\usepackage{floatflt}
\makeatletter
\renewcommand{\section}{\@startsection{section}{1}{0in}
	{0.4\baselineskip}{0.1\baselineskip}{\Large\bf}}
\renewcommand{\subsection}{\@startsection{subsection}{2}{0in}
	{0.25\baselineskip}{-\baselineskip}{\large\bf}}
\renewcommand{\subsubsection}{\@startsection{subsubsection}{3}{0in}
	{0.1\baselineskip}{-\baselineskip}{\normalsize\bf}}
\makeatother
\begin{document}

%
%  Session and Paper Code:
\thispagestyle{myheadings}
%
%  ***INSTRUCTIONS:***  Replace `OG 9.9.9' in the command argument below
%			with your assigned session and paper code:
\markright{OG 4.3.31}
\begin{center}
%
%  ***INSTRUCTIONS:***  Replace `Instructions for Preparation of Manuscript'
%			with your paper's title:
{\LARGE \bf Data Acquisition System of the CANGAROO-II Telescope}
\end{center}

%  Author List:
\begin{center}
%
%  ***INSTRUCTIONS:***  Replace authors and addresses below with your own:
%
%{\bf Masaki Mori$^{1}$ for the CANGAROO Team}\\
%{\it $^{1}$Institute for Cosmic Ray Research, University of Tokyo, Tanashi, Tokyo 188-8502, Japan}
{\bf
M.~Mori\footnote[1]{
Institute for Cosmic Ray Research, University of Tokyo,
     Tanashi, Tokyo 188-8502, Japan, 
$^{2}$Department of Physics and Mathematical Physics, University of 
   Adelaide, South Australia 5005, Australia, 
$^{3}$Institute of Space and Astronautical Science,
   Sagamihara, Kanagawa 229-8510, Japan, 
$^{4}$Department of Physics, Yamagata University, 
Yamagata 990-8560, Japan, 
$^{5}$Department of Physics, Tokyo Institute of Technology, 
        Meguro, Tokyo 152-8551, Japan, 
$^{6}$Faculty of Management Information, Yamanashi Gakuin Univeristy,  Kofu, Yamanashi 400-8575, Japan, 
$^{7}$Department of Physics, Tokai University, 
 Hiratsuka, Kanagawa 259-1292, Japan, 
$^{8}$STE Laboratory, Nagoya University,
   Nagoya, Aichi 464-860, Japan, 
$^{9}$National Astronomical Observatory, Tokyo 181-8588, Japan, 
$^{10}$Faculty of Science, Ibaraki Univeristy, 
   Mito, Ibaraki 310-8521, Japan, 
$^{11}$LPNHE, Ecole Polytechnique. Palaiseau CEDEX 91128, France,
$^{12}$Institute of Physical and Chemical Research,
   Computational Science Laboratory, Institute of Physical and Chemical
   Research, Wako, Saitama 351-0198, Japan,
$^{13}$Faculty of Engineering, Kanagawa University,
 Yokohama, Kanagawa 221-8686, Japan
}, 
S.A.~Dazeley$^2$,
P.G.~Edwards$^3$,
S.~Gunji$^4$, S.~Hara$^5$,  
T.~Hara$^6$, J.~Jinbo$^7$, 
A.~Kawachi$^1$, T.~Kifune$^1$, 
H.~Kubo$^5$, 
J.~Kushida$^5$, Y.~Matsubara$^8$, 
Y.~Mizumoto$^9$, 
M.~Moriya$^5$, 
H.~Muraishi$^{10}$, Y.~Muraki$^8$, 
T.~Naito$^6$, K.~Nishijima$^7$, 
J.R.~Patterson$^2$, M.D.~Roberts$^1$, 
G.P.~Rowell$^1$, T.~Sako$^{8,11}$, 
K.~Sakurazawa$^5$, Y.~Sato$^1$, R.~Susukita$^{12}$, 
T.~Tamura$^{13}$,  
T.~Tanimori$^5$, S.~Yanagita$^{10}$, 
T.~Yoshida$^{10}$, T.~Yoshikoshi$^1$, and
A.~Yuki$^8$  
}
\end{center}

%  Abstract:
\begin{center}
{\large \bf Abstract\\}
\end{center}
\vspace{-0.5ex}
%
%  ***INSTRUCTIONS:***  Replace text below with your own abstract:
%
The data acquisition system for the new CANGAROO-II 7m telescope is
described. 
%

%  Leave this line skip in place:
\vspace{1ex}

%
%  Manuscript text:
%
%  ***INSTRUCTIONS:***  Delete the next few pages of text and enter your own.  There will
%			be a warning, `STOP DELETING TEXT!!', just before the References
%			section so that the standardized Reference heading will not be
%			accidently erased.  Within the text below is an example is given
%			of a figure placement (using `picinpar').
\section{Introduction:}
\label{intro.sec}

The CANGAROO-II 7m telescope which has just been completed in Woomera,
Australia aims to extend our study of very-high-energy gamma-ray
astrophysics to the hundreds GeV region.

The mechanical and optical design of the telescope are 
reported in Tanimori et al.\ 1999 and Kawachi et al.\ 1999, respectively: 
here we describe the data acquisition system. 
% which is shown schematically in Figure \ref{fig:daq}.

%\begin{figure}
%\begin{center}
%\label{fig:daq}
%\epsfig{file=C-ii-sys.eps}
%\caption{A scheme of the data acuisition system of the CANGAROO-II telescope.}
%\end{center}
%\end{figure}

\section{Hardware:}
\label{desc.sec}

\begin{figure}[h]
\begin{center}
\label{fig:ampmod}
\epsfig{file=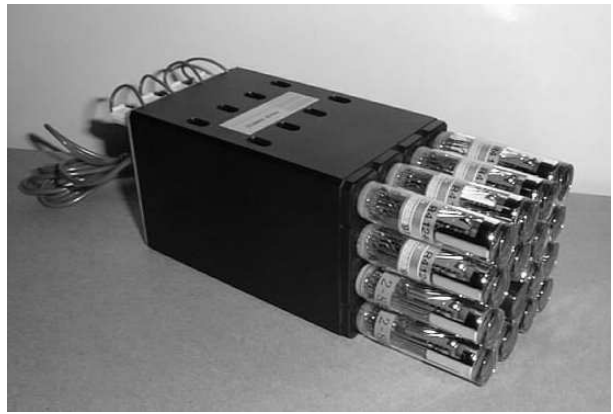, width=8cm}
\caption{A PMT base and amplifier module with sixteen 13mm photomultipliers. 
The PMT-base face size is 64mm$\times$64mm.}
\end{center}
\end{figure}

\begin{figure}
\begin{center}
\label{fig:block}
\epsfig{file=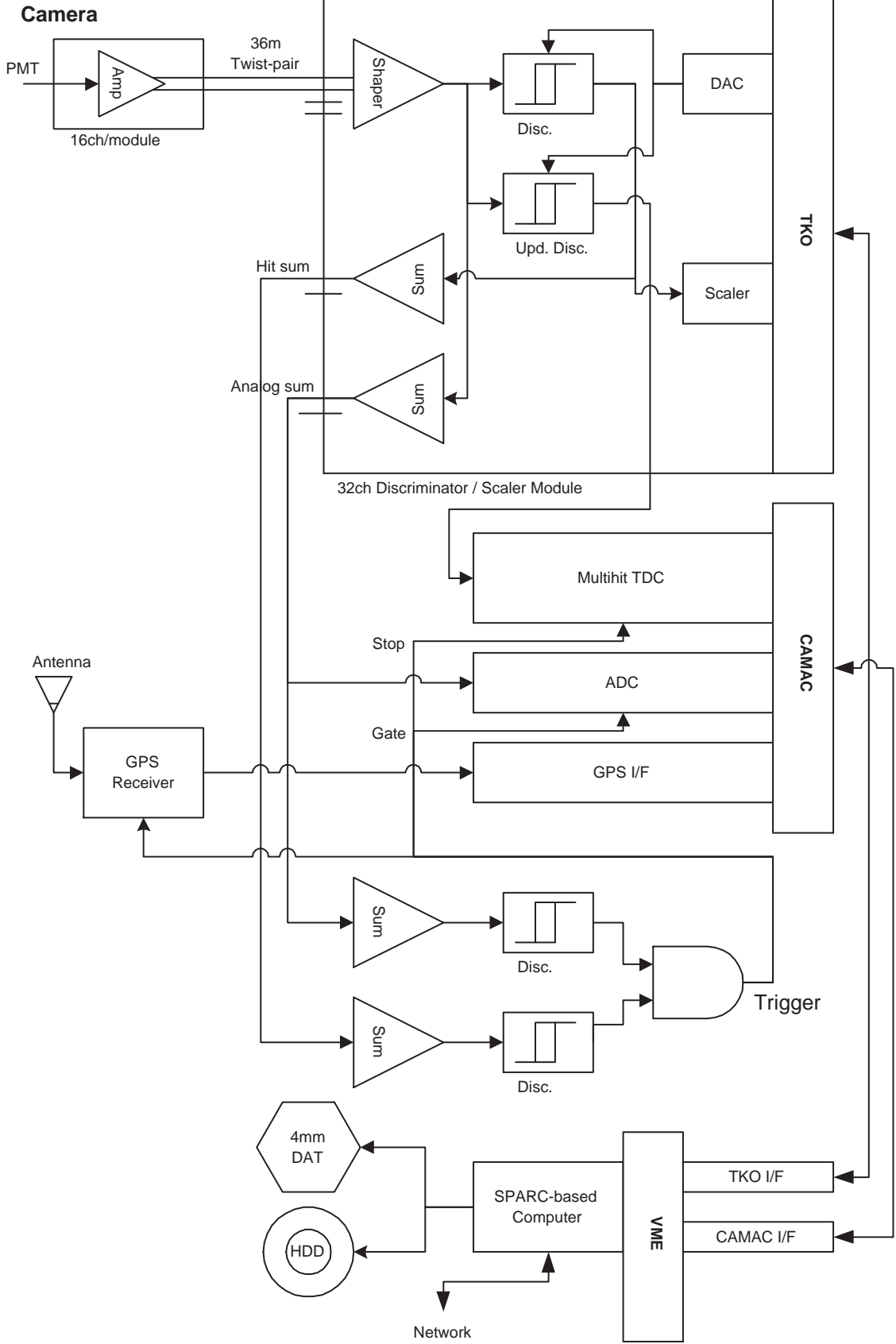}
\caption{A block diagram of the electronics of the CANGAROO-II telescope.}
\end{center}
\end{figure}

\subsection{Camera}
The prime focus of the reflector is equipped with a multi-pixel camera 
consisting of 512 fast photomultipliers (PMTs) of 13mm diameter with
UV glass windows (Hamamatsu R4124UV) supplemented by light collecting cones
which reduce the dead space between photosensitive area of the PMTs.
A PMT base and amplifier module which houses 16 photomultipliers 
%is shown in Figure \ref{fig:ampmod}.
is shown in Figure 1.
A total of 32 modules arranged in 6$\times$6 square with the four
``corner" modules missing
make up the camera, which subtends a field-of-view of about 3 degrees.
The signal from each PMT is fed into a preamplifier of gain $\sim100$ 
next to the PMT base and is transmitted through twist-pair cables of
36m in length to the electronics modules located inside the
electronics hut.
The risetime of signals is about 5ns after the preamplifier
and about 15ns after the twisted-pair cable.

\subsection{High voltage}
A high voltage power supply (LeCroy 1454/1461) gives a bias voltage
to each PMT to obtain a rather low gain of $10^5$ in order to avoid
possible overcurrent. The voltage is controlled and monitored 
via an RS232-C line by a host computer and adjusted and monitored.
All 16 PMTs in the same base and amplifier module have the same
high voltage, and so the PMTs for each module are carefully
selected to ensure they have similar gains.

\subsection{Signal processing electronics}

%The block diagram is shown in Figure \ref{fig:block}.
The block diagram is shown in Figure 2.

\begin{figwindow}[2,r,%
%\label{fig:curve}
%{\mbox{\epsfig{file=Result_16chs.eps, width=10cm}}},
{\mbox{\epsfig{file=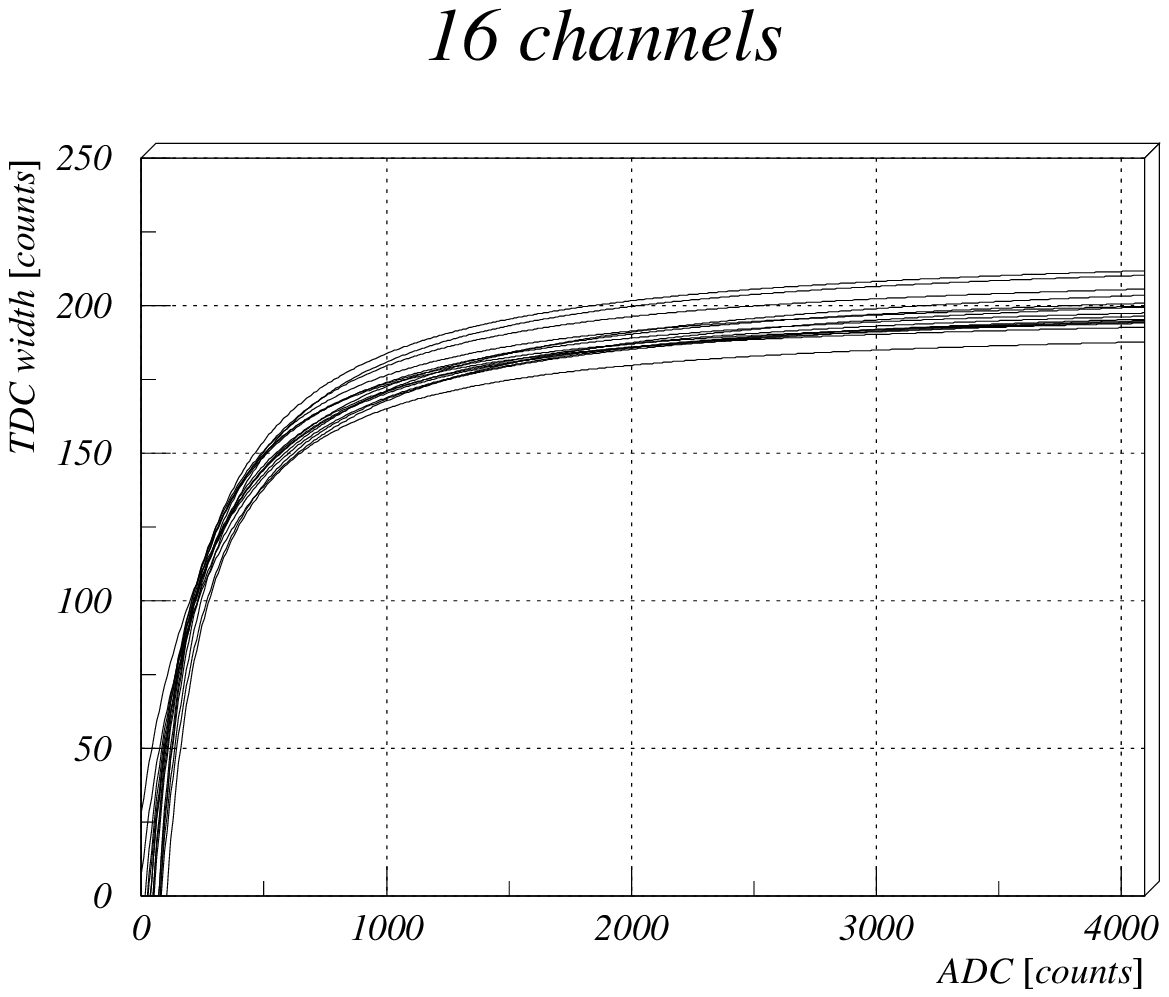, width=11cm}}},
{An example of the pulse height vs.\ pulse width (or time over
threshold) curve. Sixteen curves from 16 PMTs assmbled in a module
(Figure 1) are shown.
These are measured in a laboratory with the same electronics and cables
using an LED light source.}]
The discriminator and scaler module (DSM) works as a front-end
custom-made module based on the TKO specification (Ohska et al.\ 1985).
The signal from each PMT is shaped first and then fed into two discriminators:
a non-updating one to form event triggers and the other, an
updating one, to be sent to a multihit TDC (LeCroy 3377).
These discriminator levels can be adjusted on-line via the TKO bus.
Also included in this module are two summed outputs:
one is a linear sum of 16 input signals and the other is proportional
 to the number of hit channels out of 16 in the module 
 (``hitsum").
Two sets of these circuits are assembled in one board and make
up a 32 channel module.

The signal width, or time over threshold, measured with 0.5ns resolution 
in the TDC, gives information
%of the pulse height in a logarithmic way (Figure \ref{fig:curve}).
of the pulse height in a logarithmic way (Figure 3).

The linearly summed signal is sent to a CAMAC-based ADC (CAEN C205)
and also used for event triggers.
The former information supplements the signal amplitude in unit of
the PMT base and amplifier module.

The 12-bit scalers in the DSM are activated for a short time
(adjustable internally between 100$\mu$s and 1ms)
 by an external trigger to provide
a measure of the night sky background.
\end{figwindow}

%\begin{figure}
%\begin{center}
%\label{fig:curve}
%\epsfig{file=Result_16chs.eps, width=10cm}
%\caption{An example of the pulse height vs.\ pulse width (or time over
%threshold) curve. Sixteen curves from 16 PMTs assmbled in a module
%(Figure 1) are shown.
%These are measured in a laboratory with the same electronics and cables
%using an LED light source.}
%\end{center}
%\end{figure}

\subsection{Trigger}
Linear and {\it hit-}sums from 16 DSMs are summed in a linear adder
module and fed into discriminators.
An event is triggered when both of these summed signals 
become larger than preset values.
This starts the data acquisition sequence via an interrupt
register module on a CAMAC dataway.

\subsection{Data readout}

A VME-based board computer (FORCE CPU-7V, TurboSPARC 170MHz),
 running the Solaris
2.6 operating system and the UNIDAQ software (Nomachi et al.\ 1994), 
reads data from TDCs
and an ADC through a VME/K-bus interface (Kinetic 2917) and 
a CAMAC crate controller (Kinetic 3922) upon receipt of a trigger.
The multihit TDC records all leading and trailing edges of signals
 during 512ns with a double hit resolution of 10ns: 
 this feature can also be used to cut and monitor the night
sky background, since the time range is far larger than the Cherenkov
light pulse duration.

Presently the system dead time is about 15\% and 30\% when
the trigger rate is 12Hz and 28Hz, respectively.

PMT counting rates measured in the DSMs are read out with the 1 PPS
(pulse per second)
signals from a GPS time watcher (System Arts SA-870) and 
used to monitor bright stars.

Finally the data is stored on the local hard disk 
and is moved to a 4mm DAT
after observation for off-line analysis.

\subsection{Tracking}
The alt-azimuth mount of the telescope is driven by a telescope
controller based on a 68K processor board which accepts 
coordinates of the telescope in azimuth and elevation.
The tracking computer calculates the designated coordinates
and sends the values via an RS232-C line every 100ms. 
In parallel, encoder values are received.
A real-time Linux (KURT) operating system has been adopted
to achieve fast response time while retaining full 
Unix environment capability.
The computer clock is synchronized
to UTC utilizing the NTP software with a GPS receiver (Furuno TS-800)
or world-wide time servers through dial-up internet connection
in order to maintain tracking accuracy with errors less than 1 arcmin.

\subsection{Calibration}
At present we use a blue LED (Nichia Chemical) set at the center of the
reflector and driven by a fast pulser as a light source 
for field-flattening of the camera. 
The use of laser pulses for faster timing is under consideration.

Relative event arrival times are measured with a CAMAC scaler counting 1 MHz
clock pulses. The absolute timing of this clock is calibrated 
using the GPS time stamps
encoded at a lower rate than the event trigger, due to the limited
transferring rate of the GPS time receiver.
This global timing system will be soon replaced by a faster VME-based
GPS receiver module which gives time stamps to all events.

Telescope tracking accuracy was measured by observing stars at various
elevations and azimuths
with a CCD camera (SBIG ST-7) and a telescopic lens placed at the center of
the reflector before installing the camera.
The result shows the errors are well below 1 arcmin irrelevant to
telescope positions, even at low elevations.

\subsection{Monitor}
Events are monitored in near real time, locally or via network,
through an event display program utilizing
the buffer managing capability of the UNIDAQ software.
The voltages and currents supplied to PMTs are measured regularly 
and their values are logged.
The sky condition is also monitored by a CCD camera (SBIG ST-5C)
placed next to the camera and covering a similar field-of-view to the
PMT camera.

The telescope tracking data, consisting of designated coordinates 
to the telescope controller and coordinates read from encoders,
are regularly sent to the data-taking computer through network
and recorded with the PMT data in order to ensure the telescope
is pointing the expected direction.

\section{Summary:}
The data acquisition system for the CANGAROO-II 7m telescope
has begun to collect data to investigate astrophysical gamma-rays 
in the several hundred-GeV region. 
We are planning to install ADCs to
extend the dynamic range of observable gamma-ray energies.
The data aquisition system for CANGAROO-III, 
an array of four 10m telescopes, 
will be designed with the experience gained during these developments.

%
%  References: (DO NOT ALTER NEXT 4 LINES)
\vspace{1ex}
\begin{center}
{\Large\bf References}
\end{center}
%
%  ***INSTRUCTIONS:***  Enter your references alphabetically following the format
%			of the example citations below.
Kawachi, A.\ et al., 1999, Proc.\ 26th ICRC (Salt Lake City, 1999).\\
Nomachi, M.\ et al.,, 1994, Proc.\ Int.\ Conf.\ on Computing in High Energy
Physics '94, LBL-35822, pp.\ 114-116.
Ohska, T.\ K.\ et al., 1985, KEK Report 85-10 (KEK, Tsukuba, Japan).\\
Tanimori, T.\ et al., 1999, Proc.\ 26th ICRC (Salt Lake City, 1999).\\

\end{document}